\documentclass[10pt,draft]{dis03}
\usepackage{epsf,amsmath}

\begin{document}

\title{NNLO parton distributions from deep-inelastic-scattering data}

\author{S.~Alekhin \\IHEP, Protvino \\E-mail: alekhin@sirius.ihep.su}

\maketitle

\begin{abstract}
\noindent
The parton distributions functions (PDFs) derived
from the NNLO QCD analysis of existing light-targets
deep-inelastic-scattering data are presented. The NLO and NNLO PDFs
are compared in order to analyze perturbative stability
of the analysis and estimate impact of the higher-order QCD
corrections. The main theoretical uncertainties and experimental uncertainties
in PDFs due to all sources of experimental errors in data
are estimated and used to assess corresponding
uncertainties in the cross sections of other hadronic processes.
\end{abstract}

A study of the hard processes in the hadron collisions  
requires knowledge of the parton distribution functions (PDFs). The 
PDFs determine the normalization of the cross sections 
and therefore very often the uncertainty in the PDFs puts limit on 
the precision of measurements. For this reason the estimation 
of the PDFs uncertainties and improvement of our knowledge of the PDFs
was the scope of intensive studies in recent years~\cite{Giele:2002hx}.
Estimation of the different uncertainties in PDFs shows that the 
uncertainties in the ansatz of analysis (theoretical ones)
dominate over the uncertainties 
propagated from the experimental errors in data used to constrain
the PDFs (experimental ones) for wide kinematics~\cite{Botje:1999dj}. 
One of the potentially 
dangerous theoretical uncertainty in the PDFs extracted in the NLO QCD 
comes from the higher orders (NNLO) corrections. The NNLO
corrections are unknown for the most of hard processes, but 
have been almost completely known for the DIS
with calculation of the splitting function moments up to 
the 12-th~\cite{Retey:2000nq}.
The remaining uncertainty in the DIS structure functions due to 
missing terms in the NNLO correction 
was estimated at the level of few percent 
for the realistic kinematics~\cite{vanNeerven:2000wp}.

\begin{figure}[h]
\vspace*{7.0cm}
\begin{center}
\includegraphics{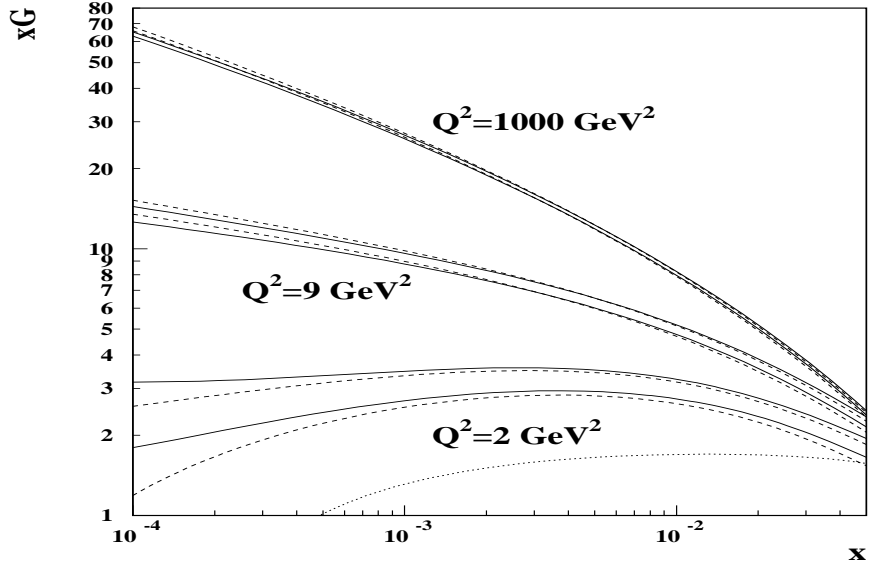}
\vskip 0.5cm
\caption[*]{The $1\sigma$ experimental error bands
for the A02 gluon distributions obtained
in the NNLO (solid lines) and the NLO (dashes) at different values of $Q$.
The NNLO gluon distribution for Set 1 of MRST2001 at 
$Q^2=2~{\rm GeV}^2$ is given for comparison (dots).}
\end{center}
\end{figure}

The approximate NNLO corrections of Ref.~\cite{vanNeerven:2000wp} were  
applied
in the QCD analysis of the existing charged-leptons DIS data including 
data from the SLAC, BCDMS, NMC, H1, and  ZEUS 
experiments~\cite{Alekhin:2002fv}.
The NNLO and the NLO gluon distributions obtained in this analysis (A02) are 
compared in Fig.1. Variation of the gluon distribution 
under account of the NNLO correction is generally within the experimental 
errors that indicates perturbative stability of NNLO PDFs. 
In the same plot we give the NNLO gluon distribution extracted from the 
MRST global fit of ~Ref.\cite{Martin:2002dr}.
At low $Q$ the MRST gluon distribution is well below the A02 one and gets 
negative at $x \sim 0.0001$. The origin of this difference
has not been clarified entirely. One of the possible reasons  
is certain inconsistency of the NNLO MRST fit due to  
the NNLO corrections have been applied for the DIS only, while the 
MRST fit 
includes the Drell-Yan (DY) and the jet production data as well.
This inconsistency was explicitly demonstrated in study of impact of the 
NNLO correction on the DY cross sections~\cite{Anastasiou:2003yy}:
The NLO DY cross sections calculated using MRST PDFs are in good 
agreement with experimental data, while the NNLO predictions are 
well above the data. Note for the jet data used in the MRST analysis the 
NNLO corrections are presumably even more important than for the DY process.
Therefore, for the NNLO
PDFs extracted from the DIS data the higher-order QCD
theoretical 
errors are under much better control than for PDFs from the global fits.

\begin{figure}[h]
\vspace*{7.0cm}
\begin{center}
\includegraphics{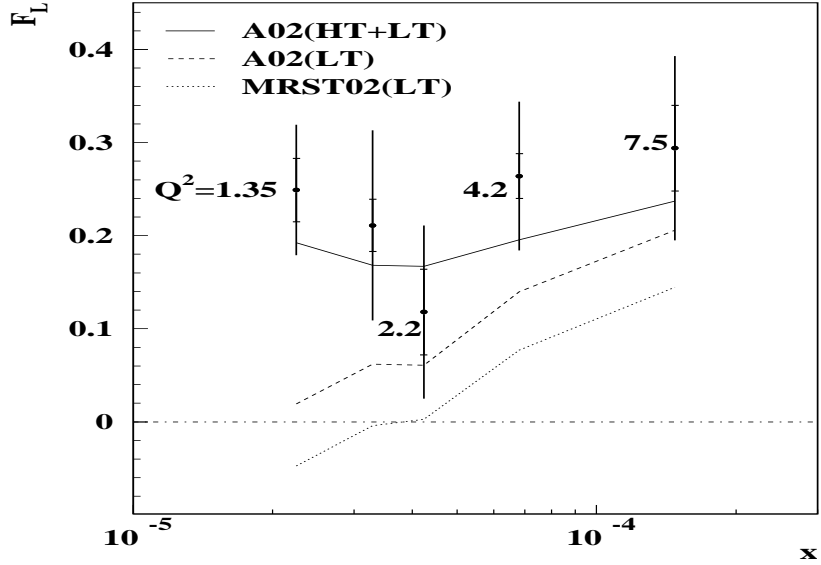}
\vskip 1cm
\caption[*]{The measurements of $F_{\rm L}$ 
by the H1 collaboration~\protect\cite{H1}
compared to the NNLO QCD predictions (solid lines: 
A02(leading twist+high twist); dashes: A02(leading twist); dots: MRST(leading
twist)).}
\end{center}
\end{figure}

The NNLO QCD predictions for the structure function $F_{\rm L}$ based 
on the different sets of PDFs are compared to the H1 measurements 
of Ref.\cite{H1} in Fig.2. The MRST predictions are everywhere 
below the data. 
Since for the kinematics of Fig.2 $F_{\rm L}$ is defined mainly 
by the gluon distribution this might mean that the 
MRST underscore the gluon distribution at small $x$ and $Q$.
For the A02 prediction agreement with the data is much better.
At $x \sim 0.0001$ the agreement is achieved mainly due to the leading-twist 
(LT) term, which was calculated using the NNLO PDFs.
For $x \sim 0.00001$ the main contribution comes from the high-twist (HT)
term in $F_{\rm L}$, which was fitted in the A02 analysis 
simultaneously with the leading-twist term.
This HT term might be manifestation of a new phenomena 
expected at small $x$ or can be caused by the conventional 
effects like missing resummation of large logs
or, eventually, by a disagreement of
data from separate experiments used in the combined fit. 
In any case the unusual behavior of $F_{\rm L}$ at small $x$ and $Q$
must stimulate more precise measurements in this region.

\begin{figure}[t]
\vspace*{7.0cm}
\begin{center}
\includegraphics{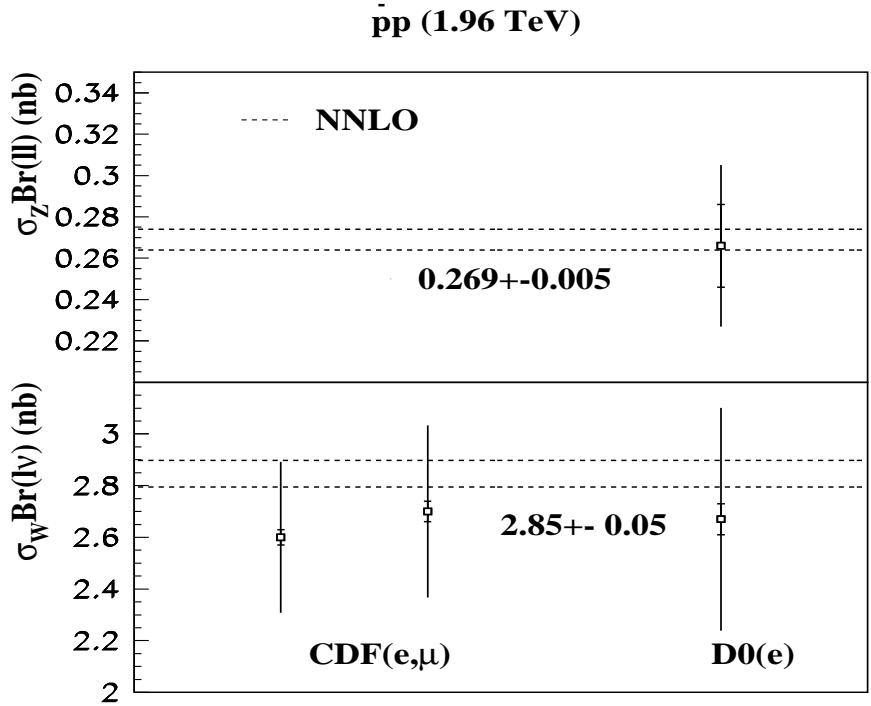}
\vskip 1.7cm
\caption[*]{The NNLO $W/Z$ production rates in
the $\overline{p}p$ collisions at $\sqrt{s}=1.96~{\rm TeV}$ compared to the
preliminary results for Run II \protect\cite{Evans:2002wj}. 
The area between dashes gives $1\sigma$ band
uncertainty in the calculations due to theoretical and experimental 
uncertainties in the PDFs.}
\end{center}
\end{figure}

With a good control of theoretical errors the NNLO PDFs  
extracted from the inclusive DIS data provide good 
tool for the precision measurements on the hadron colliders.
The NNLO predictions for the $W/Z$ cross sections
obtained using the code of Ref.\cite{Hamberg:1990np}
with the A02 PDFs are compared to the experimental data 
in Fig.3. The prediction
error bands given in Fig.3 includes both theoretical and 
experimental errors in PDFs. The accuracy of predictions is much better 
than the errors in data, which are mostly due to uncertainty in 
the luminosity. Therefore these prediction combined with the measurements 
of the $W/Z$ rates can be used for better monitoring 
of the rates of other processes including exotic ones. 

\section*{Acknowledgments} The work was supported by the RFBR grant 
03-02-17177.

\end{document}